\documentclass[conference]{IEEEtran}

\usepackage{amsmath,amssymb}
\usepackage{bm}
\usepackage{graphicx}
\usepackage{booktabs}
\usepackage{cite}
\usepackage{microtype}
\usepackage{flushend}

\newcommand{\N}{\mathbb{N}}
\newcommand{\given}{\,|\,}
\DeclareMathOperator*{\argmin}{arg\,min}
\DeclareMathOperator*{\argmax}{arg\,max}

\title{Exact Joint MAP Decoding for Unknown Load and Message Counts in Orthogonal Unsourced Random Access}

\author{\IEEEauthorblockN{Mohammad Kazemi\IEEEauthorrefmark{1} and Mert Ozates\IEEEauthorrefmark{2}}
\IEEEauthorblockA{\IEEEauthorrefmark{1}Department of Computing, Imperial College London, London, UK\\
\IEEEauthorrefmark{2}IHP -- Leibniz Institute for High Performance Microelectronics, Frankfurt (Oder), Germany}}

\begin{document}
\maketitle

\begin{abstract}
We study joint recovery of the load (the number of active users) and message-count vector in type-based unsourced random access with very short messages.  Each user's $k$-bit message activates one of $M=2^k$ orthogonal resources; coherent collisions therefore produce a noisy, unnormalized histogram of the transmitted messages.  For arbitrary load and message priors, we propose an exact joint maximum-a-posteriori (MAP) decoder.  Known-load allocation is separable and convex; our contribution is to show that the solutions for all candidate loads are nested prefixes of one greedy chain.  Scoring these prefixes with the Bayesian posterior terms jointly estimates the load and message-count vector in $O(M+K_{\max}\log M)$ time.  We compare the proposed joint MAP decoder with marginal MAP, three scalar-count two-stage receivers, oracle-load MAP, a sum-only receiver, and a soft-pilot receiver.  For $K\sim\operatorname{Unif}\{0{:}60\}$ at 2 dB, the proposed decoder cuts the no-activity false-alarm probability from $25.9\%$ to $3.5\%$ relative to the Bayesian-count two-stage receiver and, conditioned on $K\ge1$, reduces normalized count error by $21.4\%$.  The advantage concentrates under idle or sparse uncertain activity, whereas support-count processing approaches joint MAP when the possible loads are well separated.  We also prove that the sum-only statistic is insufficient and quantify prior and amplitude mismatch.  The model targets synchronized, power-controlled, very-short payloads rather than general URA.
\end{abstract}


\section{Introduction}
Unsourced random access (URA) asks a receiver to recover an unordered collection of short messages from an unknown active set~\cite{polyanskiy2017perspective,ozates2026survey}.  Ordinary URA often reports a repeated message only once.  In event reporting, however, its multiplicity is itself useful: a roadside unit may need the number of vehicles reporting each hazard or road state.  This message-count target connects URA to type-based estimation~\cite{mergen2006type} and type-based unsourced multiple access (TUMA)~\cite{ngo2024tuma,krishnan2025bound}.

We isolate the unknown-load question in a simple short-message regime.  A $k$-bit message activates one of $M=2^k$ real orthogonal resources, producing one bin of the received histogram.  Exponential blocklength and high peak-to-average power ratio preclude a general URA claim, but payloads near ten bits give a useful model for low-duty-cycle, synchronized devices.

A recent concurrent work gives a greedy maximum-likelihood (ML) detector for type-based over-the-air computation when the total count is known~\cite{martinez2026ml}.  Greedy fixed-cardinality allocation is also a classical separable resource-allocation result~\cite{fox1966marginal,ibaraki1988resource}.  We do not claim that component as new.  Our contribution is the unknown-load Bayesian extension: arbitrary message probabilities preserve increasing marginal costs; all fixed-load optima are nested in one chain; and load-prior and multinomial-mass terms select the joint MAP solution in one greedy-chain pass.

We further (i) prove why a sum-only activity statistic discards information; (ii) compare joint MAP with independent marginal MAP, three scalar-count two-stage receivers, oracle-load MAP, and a joint soft-pilot receiver under the same energy; and (iii) characterize how performance depends on load-prior geometry and receiver mismatch.  Unlike a hard pilot decision, the pilot receiver fuses pilot and data likelihoods and reduces to joint MAP at zero pilot energy.  We include hotspot traffic to exercise multiplicities, ordinary binomial activity, prior-support mismatch, and user-level amplitude jitter.

\subsection{Positioning relative to TUMA and OTA}
The recovery target here differs from three related formulations.  A conventional URA list records which messages occurred and may discard repeated instances.  TUMA instead seeks the empirical message distribution, so retaining counts is essential; existing TUMA work develops coding formulations and achievability bounds for general channels~\cite{ngo2024tuma,krishnan2025bound}.  Over-the-air computation (OTA) also exploits waveform superposition, but usually targets a prescribed low-dimensional function such as a sum or average.  Here the output is the full $M$-dimensional unnormalized histogram together with its previously unknown total mass.  Accordingly, the sum of all bins is one noisy activity feature, not the computation target.

Our scope is narrower than general TUMA but gives a complete receiver result for the orthogonal pulse codebook.  When $K$ is supplied, minimizing the Gaussian fitting cost over integer message-count vectors is the allocation problem studied in~\cite{martinez2026ml}; its greedy/SIC solution overlaps the fixed-cardinality part of Theorem~1.  Unknown $K$ changes the decision problem: the multinomial coefficient, nonuniform $\bm p$, and load prior $\bm\pi$ all enter the posterior.  The key structural fact is that these additions preserve the ordering within each marginal-cost chain, while the $q$-only terms can be evaluated on its nested prefixes.  Thus joint MAP needs neither a separately designed activity rule nor repetition of the known-load solver over all $q$.  This is a structural and computational claim: a well-designed two-stage receiver can be nearly equivalent when its initial count resolves widely separated load hypotheses, but not under every prior.

\section{Model and Joint MAP Objective}
Let the load $K$, i.e., the number of active users, take values in $\{0,\ldots,K_{\max}\}$ with prior $\pi_q$, and let each active user independently select message $m$ with probability $p_m>0$.  The message-count vector obeys
\begin{equation}
 \bm N=(N_1,\ldots,N_M)^\mathsf{T},\qquad
 \bm N\given K=q\sim\operatorname{Multinomial}(q,\bm p).
\end{equation}
Here $\bm N$ is the message-count vector, or unnormalized message histogram, and $N_m$ is the multiplicity of message $m$.  For $K>0$, the normalized empirical type is $\bm\tau=\bm N/K$; unlike $\bm N$, it does not retain the load.
With coherent equal-amplitude pulses,
\begin{equation}\label{eq:model}
 \bm Y=a\bm N+\bm Z,\qquad \bm Z\sim\mathcal N(\bm0,\sigma^2\bm I_M).
\end{equation}
For $M=2^k$, $E_b/N_0=a^2/(2k\sigma^2)$ under the real-noise convention $N_0=2\sigma^2$.  Up to constants independent of $\bm n$, the negative log-posterior is
\begin{align}
 J(\bm n)=&\sum_{m=1}^{M}\left[
 \frac{(y_m-a n_m)^2}{2\sigma^2}+\log(n_m!)-n_m\log p_m\right]\nonumber\\
 &-\log \pi_q-\log(q!),\qquad q=\|\bm n\|_1. \label{eq:objective}
\end{align}
The joint MAP estimate minimizes~\eqref{eq:objective} over $\bm n\in\N_0^M$ with $\|\bm n\|_1\le K_{\max}$.  Direct enumeration has $\sum_q\binom{M+q-1}{q}$ candidates.  The nonuniform message prior changes the binwise costs, whereas $-\log\pi_q-\log(q!)$ scores the total load; Section~III shows how one nested chain handles both effects.

\section{One Greedy Chain Solves Every Load}
For fixed $q$, the last two terms of~\eqref{eq:objective} are constant.  The incremental cost of changing coordinate $m$ from $r$ to $r+1$ is
\begin{equation}\label{eq:increment}
 \Delta_m(r)=\frac{a^2(2r+1)-2ay_m}{2\sigma^2}
 +\log(r+1)-\log p_m.
\end{equation}
It is strictly increasing since
\begin{equation}\label{eq:convex}
 \Delta_m(r+1)-\Delta_m(r)=\frac{a^2}{\sigma^2}
 +\log\frac{r+2}{r+1}>0.
\end{equation}

Intuitively, each bin offers a queue of increasingly costly additional users: repeatedly taking the cheapest available increment produces one sequence whose length-$q$ prefix is optimal for every candidate load $q$.

\textbf{Theorem 1:} Sort implicitly the $M$ increasing chains
$\{\Delta_m(r):r\ge0\}$.  For each $q\le K_{\max}$, selecting the $q$ smallest increments yields a fixed-load optimum.  These optima are nested; the prefix minimizing the complete score~\eqref{eq:objective} is the joint MAP estimate.

\emph{Proof:} Consider only the first $q$ increments of each chain, since no later increment can enter a size-$q$ allocation.  By~\eqref{eq:convex}, if $\Delta_m(r)$ is selected among the $q$ smallest, every predecessor $\Delta_m(s)$, $s<r$, is smaller and is also selected.  Thus, the $q$ smallest increments automatically form a feasible union of coordinate prefixes.  Any other feasible size-$q$ selection replaces at least one of them by an increment no smaller, so its separable cost cannot be lower.  Adding the next globally smallest available increment constructs the size-$(q+1)$ optimum without changing the first $q$ selections.  Finally, for each prefix add the fixed-cardinality term $-\log\pi_q-\log(q!)$ and choose the lowest score. \hfill$\square$

This is an exchange argument for separable convex allocation, in the spirit of marginal analysis~\cite{fox1966marginal,ibaraki1988resource}; the new step here is scoring all nested cardinalities under the load and multinomial priors.  Implementation uses a min-heap initialized with $\Delta_m(0)$.  Each pop increments one coordinate and inserts its next marginal cost.  Recording the chosen coordinate and full score after each of $K_{\max}$ pops costs $O(M+K_{\max}\log M)$ time and $O(M+K_{\max})$ memory.  After selecting $\widehat K$, replaying its prefix reconstructs $\widehat{\bm N}$.  Exhaustive enumeration over 500 random small instances gave zero objective gap to numerical precision.

\begin{figure}[t]
\centering
\footnotesize
\begin{minipage}{0.95\columnwidth}
\hrule\vspace{2pt}
\textbf{Algorithm 1: Exact joint MAP decoder}
\vspace{2pt}\hrule\vspace{2pt}
\begin{enumerate}
\setlength{\itemsep}{0pt}\setlength{\parskip}{0pt}
\item Set $\bm n=\bm0$, $C_0=\sum_m y_m^2/(2\sigma^2)$; initialize a min-heap with $(\Delta_m(0),m)$ for every $m$.
\item For $q=1,\ldots,K_{\max}$, pop $(\delta,m)$, record $m_q=m$, set $n_m\leftarrow n_m+1$ and $C_q=C_{q-1}+\delta$, then insert $(\Delta_m(n_m),m)$.
\item Choose $\widehat K=\argmin_q\{C_q-\log\pi_q-\log(q!)\}$, skipping hypotheses with $\pi_q=0$.
\item Replay $m_1,\ldots,m_{\widehat K}$ from zero counts and return the resulting $\widehat{\bm N}$.
\end{enumerate}
\vspace{1pt}\hrule
\end{minipage}
\end{figure}

\section{What a Sum-Only Receiver Discards}
The sum-only statistic $T=\bm1^\mathsf{T}\bm Y$ satisfies
\begin{equation}\label{eq:scalar}
 T=aK+W,\qquad W\sim\mathcal N(0,M\sigma^2).
\end{equation}
It is therefore an exact scalar load channel, but it adds noise from all empty bins.  Let $\bm P=\bm I-\bm1\bm1^\mathsf{T}/M$ and $\bm R=\bm P\bm Y$.  Conditional on $K$,
\begin{align}
 \mathbb E[\bm R\mid K]&=aK\bm P\bm p,\label{eq:rmean}\\
 \operatorname{Cov}(\bm R\mid K)&=\sigma^2\bm P+a^2K\bm P
 [\operatorname{Diag}(\bm p)-\bm p\bm p^\mathsf{T}]\bm P.\label{eq:rcov}
\end{align}

\textbf{Proposition 1:} If at least two loads have positive prior probability, $M\ge2$, $a>0$, and $\bm p$ is nondegenerate, then $T$ is not sufficient for $K$.

\emph{Proof:} Decompose the Gaussian noise into $W=\bm1^\mathsf{T}\bm Z$ and $\bm P\bm Z$.  Their cross-covariance is zero, so they are independent.  Moreover, $T=aK+W$, while $\bm R=a\bm P\bm N+\bm P\bm Z$; hence $T\perp\bm R\mid K$.  If $T$ were sufficient, it would also imply $K\perp\bm R\mid T$.  Combining the two conditional independences would make the law of $\bm R\mid K$ independent of $K$, contradicting~\eqref{eq:rmean}--\eqref{eq:rcov}.  For uniform $\bm p$, the mean vanishes but the covariance is $(\sigma^2+a^2K/M)\bm P$ on the orthogonal subspace. \hfill$\square$

Thus, full-vector Bayes load risk is no greater than sum-based risk and can be strictly smaller.  The sum is retained below as a diagnostic separated receiver, not as the strongest competitor.

For two equiprobable loads separated by $\Delta K$, the scalar MAP error is $Q(d_T/2)$, where
\begin{equation}\label{eq:separation}
 d_T=\frac{a\Delta K}{\sigma\sqrt M}
 =\Delta K\sqrt{\frac{2k(E_b/N_0)}{M}}.
\end{equation}
The $\sqrt M$ noise penalty makes close loads difficult for the sum even though sparse pulses remain individually visible.

\begin{figure*}[t]
 \centering
 \includegraphics[width=0.94\textwidth]{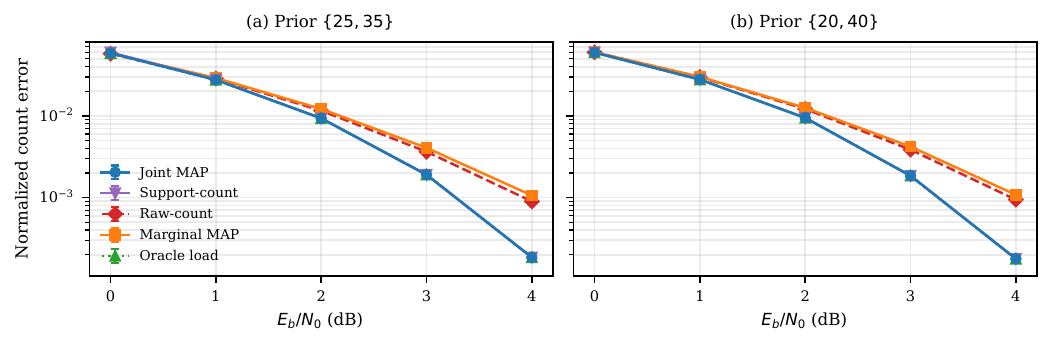}
 \caption{Two-point load priors with $10^5$ common-random-number trials per point and 95\% bootstrap intervals.  The support-count two-stage receiver nearly reaches joint MAP by snapping the marginal pulse count to the prior support; the raw-count receiver does not.}
 \label{fig:snr}
\end{figure*}

\section{Numerical Evidence}
\subsection{Receivers, metrics, and protocol}
We use $k=10$, $M=1024$, $\sigma^2=1$, and the specified $E_b/N_0$.  Marginal MAP independently estimates each $N_m$ using
\begin{equation}
 \Pr[N_m=r]=\sum_{q=r}^{K_{\max}}\pi_q\binom{q}{r}p_m^r(1-p_m)^{q-r}.
\end{equation}
Table~\ref{tab:receivers} summarizes the proposed decoder and all comparison receivers.  The three two-stage receivers first estimate the load from the marginal MAP bin decisions and then apply known-load MAP at that estimate.  The raw-count receiver uses $\widehat K_{\rm raw}=\sum_m\widehat N_m^{\rm mar}$; the support-count receiver snaps this value to
\begin{equation}
 \widehat K_{\rm sup}=\argmin_{q:\pi_q>0}|q-\widehat K_{\rm raw}|,
\end{equation}
breaking ties by larger $\pi_q$ and then smaller $q$.  The Bayesian-count receiver instead calibrates its scalar count channel on data independent of evaluation.  With $R=\sum_m\widehat N_m^{\rm mar}$, it estimates $\widehat P(R=r\mid K=q)$ from 5000 trials per supported load, using Jeffreys smoothing, and selects

\begin{equation}
 \widehat K_{\rm Bayes}=\argmax_{q:\pi_q>0}
 \pi_q\widehat P(R\mid K=q).
\end{equation}
This prior-specific receiver has comparable online order but needs offline calibration.

\begin{table}[t]
\centering
\caption{Receivers used in the numerical study.  Count-based and sum-only two-stage receivers apply known-load MAP after selecting $\widehat K$.}
\label{tab:receivers}
\scriptsize
\setlength{\tabcolsep}{2.0pt}
\begin{tabular}{@{}p{0.40\columnwidth}p{0.55\columnwidth}@{}}
\toprule
Receiver & Load information / role\\
\midrule
Proposed joint MAP & Full $\bm Y$, $\bm\pi$, $\bm p$ / proposed decoder\\
Marginal MAP & Per-bin prior marginal / nonjoint baseline\\
Raw-count two-stage & $R=\sum_m\widehat N_m^{\rm mar}$ / uncalibrated\\
Support-count two-stage & $R$ snapped to $\operatorname{supp}(\bm\pi)$ / support-aware\\
Bayesian-count two-stage & $\widehat P(R\mid K)$ and $\bm\pi$ / calibrated\\
Oracle-load MAP & True $K$ / known-load reference\\
Sum-only two-stage & $T=\bm1^\mathsf{T}\bm Y$ / sufficiency diagnostic\\
Soft-pilot joint MAP & Full $\bm Y$ and $U$ / equal-energy audit\\
\bottomrule
\end{tabular}
\end{table}

We call $H=\frac12\|\bm N-\widehat{\bm N}\|_1$ the unnormalized half-$\ell_1$ error.  For $K>0$, the normalized count error is $C=H/K$.  At $K=0$, we set $C=0$ if $\widehat K=0$ and $C=1$ otherwise.

When $K,\widehat K>0$, the true and estimated empirical types are $\bm\tau=\bm N/K$ and $\widehat{\bm\tau}=\widehat{\bm N}/\widehat K$; their total-variation error is $D=\frac12\|\bm\tau-\widehat{\bm\tau}\|_1$.  Because a type is undefined at zero load, we set $D=1$ when exactly one load is zero and $D=0$ when both are zero.

To measure multiplicity specifically in collided bins, conditioned on at least one collision, define
\begin{equation}
 C_{\ge2}=\frac{\sum_{m:N_m\ge2}|N_m-\widehat N_m|}{\sum_{m:N_m\ge2}N_m},
\end{equation}
and let $P_{\rm ex}$ be the probability that every true collided-bin multiplicity is exact.  Reported $C$ intervals are paired empirical-bootstrap intervals; binomial probabilities use Wilson intervals.

The fixed priors are uniform on $\{25,35\}$ ($K_{\max}=35$) or $\{20,40\}$ ($K_{\max}=40$).  ``Mix'' denotes an equal mixture of two $\operatorname{Binomial}(60,\rho)$ laws whose means equal the listed pair, with $K_{\max}=60$.  The hotspot prior places total probability $0.5$ on 16 bins and $0.5$ uniformly on the other 1008.  At $K=30$, the expected number of colliding user pairs is $0.425$ for uniform messages and $6.90$ for hotspot messages.

\begin{figure*}[t]
 \centering
 \includegraphics[width=0.96\textwidth]{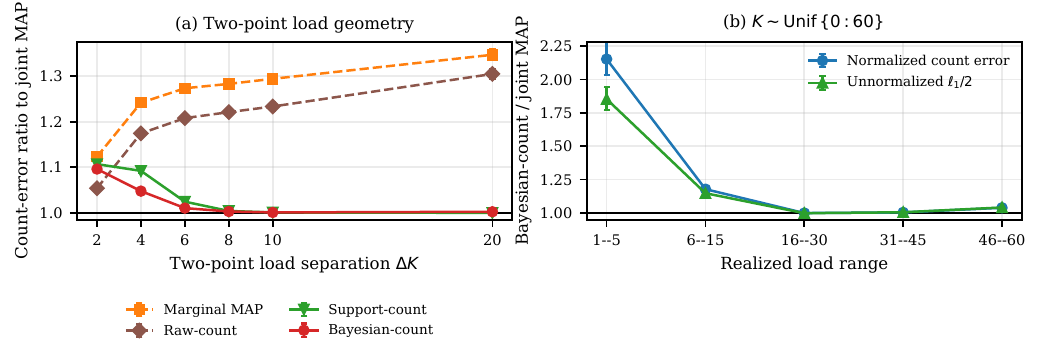}
 \caption{Load-geometry audit at 2 dB with $10^5$ held-out paired trials and 95\% bootstrap intervals.  (a) Candidate/joint count-error ratios for equiprobable two-point priors centered at 30.  (b) Bayesian-count/joint ratios under $K\sim\operatorname{Unif}\{0{:}60\}$, conditioned on $K\ge1$ and stratified by realized load; the $K=0$ result is reported separately in the text.}
 \label{fig:geometry}
\end{figure*}

\begin{table*}[t]
\centering
\caption{Results at 2 dB from $10^5$ common-random-number trials.  Entries for $C$ are mean [95\% bootstrap interval].  U/H denote uniform/hotspot messages; $P_e(K)=\Pr[\widehat K\ne K]$ for joint MAP.  The last two metrics condition on a collision event.}
\label{tab:two_db}
\scriptsize
\setlength{\tabcolsep}{2.5pt}
\resizebox{\textwidth}{!}{%
\begin{tabular}{llccccccc}
\toprule
Activity & Msg. & Joint $C$ & Support-count $C$ & Marginal $C$ & Joint $D$ & $P_e(K)$ & $C_{\ge2}$ & $P_{\rm ex}$\\
\midrule
Fixed 25/35 & U & .00934 [.00924,.00944] & .00934 [.00924,.00944] & .01216 [.01208,.01226] & .00934 & .00000 & .00666 & .98386\\
Fixed 25/35 & H & .00631 [.00622,.00640] & .00631 [.00623,.00640] & .00902 [.00895,.00910] & .00631 & .00000 & .00155 & .98574\\
Fixed 20/40 & U & .00947 [.00936,.00957] & .00947 [.00936,.00957] & .01256 [.01246,.01265] & .00947 & .00000 & .00606 & .98378\\
Fixed 20/40 & H & .00619 [.00610,.00628] & .00619 [.00610,.00628] & .00929 [.00920,.00937] & .00619 & .00000 & .00158 & .98454\\
Mix 25/35 & U & .01215 [.01206,.01224] & .01218 [.01209,.01228] & .01218 [.01209,.01227] & .02094 & .44940 & .00975 & .97602\\
Mix 20/40 & U & .01245 [.01235,.01255] & .01271 [.01261,.01281] & .01271 [.01261,.01280] & .02139 & .43712 & .00853 & .97773\\
\bottomrule
\end{tabular}}
\end{table*}

\subsection{What the separated receiver changes}
For well-separated two-point priors, enforcing the load support almost eliminates the joint-versus-two-stage gap.  At 2 dB, joint MAP lowers $C$ relative to independent marginal MAP by $23.2\%$ and $24.6\%$ for the two fixed uniform priors, but the support-count receiver is only $0.064\%$ worse for $\{25,35\}$ and exactly matches joint MAP for $\{20,40\}$.  The hotspot rows show the same equality.  Thus much of the apparent gain over marginal decisions in this setting comes from enforcing the prior support.

This equivalence strengthens with SNR.  Figure~\ref{fig:snr} shows the support-count receiver reaching joint MAP at 3--4 dB under $\{25,35\}$ and from 1 dB onward under $\{20,40\}$.  Both approach oracle-load MAP, whereas the raw-count receiver and marginal MAP remain worse; at 4 dB joint MAP lowers marginal error by $82.6\%$ and $83.5\%$.

Close load hypotheses reveal a residual gap.  Figure~\ref{fig:geometry}(a) centers two equiprobable loads at 30 and sweeps their separation $\Delta K$.  At $\Delta K=2$, the raw-count receiver is the best two-stage rule at a joint-relative ratio of $1.054$; the Bayesian- and support-count ratios are $1.096$ and $1.107$.  The Bayesian load decision minimizes load-classification error, but an incorrect support decision forces a two-user jump and need not minimize $C$.  By $\Delta K\ge8$, the support-aware receivers are within $0.3\%$ of joint MAP.  Equivalence therefore depends on the load hypotheses being resolvable, whereas joint MAP requires no separate load-estimation design.

The sum-only receiver is substantially weaker.  At 2 dB its $C$ is $0.03978$ and $0.02416$ for the fixed uniform priors, versus $0.00934$ and $0.00947$ jointly.  Equation~\eqref{eq:separation} predicts load error $0.1895$ for $\{25,35\}$, consistent with the measured $0.18791$ and with the insufficiency result.

Hotspot traffic directly exercises multiplicity recovery.  Collisions occur in $99.17\%$ and $96.70\%$ of trials, and joint MAP recovers every true collided bin exactly in $98.57\%$ and $98.45\%$ of those trials, versus $97.71\%$ and $97.60\%$ for marginal MAP.  Table~\ref{tab:dense} reports a denser diagnostic; it gives only $1.3$--$1.4\%$ total-$C$ gains over the Bayesian-count receiver and no collision-mass-error advantage, so we do not claim universal superiority in dense collisions.

\begin{table}[t]
\centering
\caption{Dense-collision diagnostic at 2 dB with $3\times10^4$ trials and $K\sim\operatorname{Unif}\{15{:}45\}$.  Entries are mean [95\% interval]; $P_{\rm ex}$ conditions on a collision.}
\label{tab:dense}
\scriptsize
\setlength{\tabcolsep}{2.0pt}
\resizebox{\columnwidth}{!}{%
\begin{tabular}{llccc}
\toprule
$M$ & Receiver & $C$ & $C_{\ge2}$ & $P_{\rm ex}$\\
\midrule
64  & Joint & .01900 [.01877,.01922] & .01801 [.01744,.01859] & .8186 [.8141,.8229]\\
64  & Bayesian-count & .01924 [.01902,.01948] & .01737 [.01680,.01791] & .8164 [.8120,.8208]\\
128 & Joint & .01756 [.01736,.01778] & .01547 [.01478,.01618] & .9016 [.8980,.9051]\\
128 & Bayesian-count & .01781 [.01760,.01803] & .01498 [.01433,.01565] & .9002 [.8966,.9037]\\
\bottomrule
\end{tabular}}
\end{table}

\subsection{Held-out soft-pilot audit}
Give a scalar pilot energy fraction $\eta$ and use
\begin{align}\label{eq:pilot}
 \bm Y&=a\sqrt{1-\eta}\,\bm N+\bm Z,\nonumber\\
 U&=a\sqrt\eta\,K+Z_p,\qquad Z_p\sim\mathcal N(0,\sigma^2).
\end{align}
The receiver adds $\ell_p(q)=(U-a\sqrt\eta q)^2/(2\sigma^2)$ to prefix $q$'s score, rather than hard-detecting $K$, and reduces to joint MAP at $\eta=0$.  Independent tuning uses $2\times10^4$ trials/point over 15 fractions from 0 to 0.3; evaluation uses a fresh $10^5$ paired trials.  Table~\ref{tab:pilot} tests random loads, where pilot-free joint MAP has $P_e(K)=0.44$--$0.45$ and an oracle gap remains.

\begin{table}[t]
\centering
\caption{Held-out equal-energy soft-pilot audit at 2 dB.  Ratios are $C_\eta/C_0$ with paired 95\% bootstrap intervals; $C_{\rm or}/C_0$ is the full-energy oracle ratio.}
\label{tab:pilot}
\scriptsize
\setlength{\tabcolsep}{2.0pt}
\resizebox{\columnwidth}{!}{%
\begin{tabular}{lccccc}
\toprule
Activity & $\eta^*$ & $C_0$ & $C_{\eta^*}$ & Ratio [95\% CI] & $C_{\rm or}/C_0$\\
\midrule
Mix 25/35 & .003 & .012216 & .012233 & 1.001 [.999,1.004] & .767\\
Mix 20/40 & .0003 & .012494 & .012503 & 1.001 [.999,1.002] & .756\\
Binom.(60,.5) & .0001 & .012012 & .012011 & 1.000 [.999,1.001] & .779\\
\bottomrule
\end{tabular}}
\end{table}

The null result has a simple energy-resolution explanation.  At 2 dB, $a/\sigma=\sqrt{20\cdot10^{0.2}}\approx5.63$.  For adjacent random loads, the pilot distance is $d_p=(a/\sigma)\sqrt\eta$: at $\eta=0.1$, $d_p=1.78$ gives pairwise error $Q(d_p/2)\approx0.187$; at $\eta=0.3$, the error remains about $0.061$ while the data lose $-10\log_{10}(0.7)=1.55$ dB.  The held-out intervals exclude improvements much above $0.1\%$.  Hence the tested scalar pilot cannot buy enough load resolution to offset its data-energy cost.

\subsection{Load geometry and mismatch}
Under a broad load prior, the advantage localizes to idle and sparse activity.  For $K\sim\operatorname{Unif}\{0{:}60\}$, the Bayesian- and support-count receivers select the same load in all $10^5$ trials, so calibration changes neither decisions nor risk.  About half of the overall count-error gap comes from $K=0$: the false-alarm probability is $0.035$ jointly and $0.259$ for either two-stage receiver.  Conditioned on $K\ge1$, joint MAP reduces $C$ by $21.4\%$ and half-$\ell_1$ error by $4.6\%$.  Figure~\ref{fig:geometry}(b) further localizes the nonidle gain: for $K=1$--5, the reductions are $53.5\%$ and $46.0\%$, respectively; they diminish for $K=6$--15 and disappear by $K=16$--30.  Thus normalization amplifies the sparse-load effect but does not create it.

The mechanism is a load-adaptive threshold.  For uniform messages, accepting a $(q+1)$st pulse in an empty bin requires
\[
 \frac{2ay-a^2}{2\sigma^2}>
 \log\frac{M}{q+1}-\log\frac{\pi_{q+1}}{\pi_q}.
\]
The threshold rises at small inferred $q$, whereas marginal MAP uses one prior-averaged scalar rule.  The tested two-stage receivers compress their initial marginal decisions to the scalar $R$ and therefore do not retain the ordered bin evidence used when joint MAP selects a prefix.  This explains their broad-prior low-load gap; it is not a claim against every possible separated architecture.

If the true load is 30 but the decoder permits only 25 or 35, then $C\ge5/(2\cdot30)$; the measured error is $0.08375$, versus $0.00938$ under a matched point mass.  Mixing $1\%$ flat mass into the two-point prior lowers it to $0.01530$, and a fully flat prior gives $0.01215$.  Full-support softening is therefore prudent when the activity model is uncertain.  With user-amplitude jitter, fixed-prior joint error grows by factors $1.078$, $1.318$, $2.269$, and $4.115$ at half-widths $0.5$, 1, 2, and 3 dB; mixture factors are $1.066$, $1.237$, $1.938$, and $3.288$.  Replacing $\sigma^2$ by a tuned $\gamma\sigma^2$, with $\gamma\in\{1,1.25,1.5,2,3,4\}$ selected on independent data, chooses $\gamma=1$ at every nonzero-jitter point and gives no held-out improvement, so simple variance inflation does not repair the mismatch.

\subsection{Scope}
The exact result assumes timing alignment, coherent equal-amplitude addition, independent Gaussian bins, i.i.d. categorical messages, and known nominal $a$, $\sigma^2$, $\bm p$, and $\bm\pi$.  Orthogonal one-pulse signaling also uses $2^k$ real dimensions and peak-to-average power ratio $M$.  The result is therefore a receiver theorem for synchronized, power-controlled, very-short payloads, not a general URA achievability claim; the support and jitter sweeps quantify its principal model sensitivities.

\section{Conclusion}
We derived the exact joint MAP decoder for the unknown load and complete message-count vector of an orthogonal pulse codebook.  The new step beyond known-load greedy allocation is that every fixed-load optimum is a prefix of one marginal-cost chain; scoring those prefixes with the load prior and multinomial mass produces the joint solution in $O(M+K_{\max}\log M)$ time.  The result applies to arbitrary load priors and nonuniform message probabilities, and it needs neither a separate activity detector nor repeated optimization for each candidate load.

The experiments identify where joint decoding matters.  The tested two-stage receivers nearly match joint MAP when a separated load support is reliably resolved, but the advantage concentrates under idle or sparse uncertain activity: for a uniform $0$--$60$ load prior, joint MAP cuts the no-activity false-alarm probability from $25.9\%$ to $3.5\%$ and, conditioned on $K\ge1$, lowers normalized count error by $21.4\%$.  Hotspot tests further show exact collided-bin recovery in $98.45$--$98.57\%$ of collision trials, about one percentage point above marginal MAP.  Dense-collision and pilot audits show no universal joint-processing gain.  Thus the contribution is an exact joint MAP decoder together with a delineation of when simpler two-stage processing is already sufficient.

\end{document}